\documentclass[iop]{emulateapj}

\usepackage{color}

\shorttitle{JPEG lensing}
\shortauthors{Lam}

\begin{document}

\title{A new approach to free-form cluster lens modeling \\ inspired by the JPEG image compression method}

\author{Daniel Lam\altaffilmark{1}}
\affil{Leiden Observatory, Leiden University, NL-2300 RA Leiden, Netherlands}
\email{daniellam@strw.leidenuniv.nl}

\begin{abstract}

I propose a new approach to free-form cluster lens modeling that is inspired by the JPEG image compression method. 
This approach is motivated specifically by the need for accurate modeling of high-magnification regions in galaxy clusters. 
Existing modeling methods may struggle in these regions due to their limited flexibility in the parametrization of the lens, even for a wide variety of free-form methods. 
This limitation especially hinders the characterization of faint galaxies at high redshifts, which have important implications for the formation of the first galaxies and even for the nature of dark matter. 
JPEG images are extremely accurate representations of their original, uncompressed counterparts but use only a fraction of number of parameters to represent that information. 
Its relevance is immediately obvious to cluster lens modeling. 
Using this technique, it is possible to construct flexible models that are capable of accurately reproducing the true mass distribution using only a small number of free parameters. 
Transferring this well-proven technology to cluster lens modeling, I demonstrate that this `JPEG parametrization' is indeed flexible enough to accurately approximate an N-body simulated cluster. 

\end{abstract}


\section{Introduction}\label{sec:intro}

One of the most active research topics on high-redshift galaxies is to measure the faint-end of their luminosity function (LF). 
The potential downturn (commonly referred to as the 'turn-over') in the number density of faint galaxies at high redshifts has important implications to many of the following questions: 
(1) what are the sources that reionized the Universe? 
(2) how do baryonic processes affect star formation efficiency in low-mass halos?
(3) what are the properties of dark matter? 

Cluster lenses provide the most promising prospect of detecting the turn-over in the faint-end LF because of their high magnification of background sources over modest areas ($\geq$ 1 arcmin$^{2}$). 
This advantage drastically reduces the investment in observational time required, compared to conventional blank, deep fields \citep{postman12, lotz17, coe19}. 
Currently, the potential turn-over in the LF is likely still a few magnitudes out of reach of even the state-of-the-art analyses of the deepest cluster-lensed deep field data \citep{atek15a, atek15b, atek18, livermore17, bouwens17, ishigaki18, kawamata18, laporte16}. 
The main challenge to measuring many valuable intrinsic properties of the lensed sources is to construct accurate lens models for magnification correction. 

From the perspective of a \textit{user} of lens models, since the underlying truth is unknown, concurrence among lens models is generally taken as an indication of their reliability. 
Conversely, where lens models show significant differences is also where users start losing confidence in the models. 
From the point of view of a \textit{modeler}, the accuracy with which various lens modeling methods recover the mass distribution of a cluster can be evaluated through end-to-end tests and modeling challenges. 
In an impressive effort, \citet{meneghetti17} performed exactly such an end-to-end test, creating a mock cluster, constructing a realistic set of data ’observing’ that mock cluster, asking various modeling groups to determine its mass distribution, and then comparing their results with the actual mass distribution. 
\citet{meneghetti17} find that the best performing method \citep[GLAFIC,][]{oguri10} is able to reconstruct the magnification of an N-body simulated cluster to within 10\% in the median at a true magnification of $\mu_{true} \approx 3$, and within 30\% in the median at $\mu_{true} \approx 10$ using strong lensing constraints. 
However, one should note that this test is the first of its kind, and so is justifiably simplistic compared to the real universe. 
For example, the mock cluster is constructed from a dark matter-only N-body simulation. 
In reality, a non-negligible fraction of the cluster mass resides in the hot intra-cluster gas, and it is not obvious that its spatial distribution should be well-described by any of the commonly chosen parametric functions (e.g. Navarro-Frenk-White (NFW) profile \citep{navarro96}), especially for merging cluster-cluster collisions, where the gas and dark matter are displaced spatially \citep{clowe06}. 

Fundamentally, magnification $\mu$ depends on the gradients of the scaled deflection field $\vec{\alpha}$, 

\begin{equation}
\mu (\vec{\theta}) = \bigg[ \mathrm{det} \bigg( \delta_{ij} - \frac{\partial \alpha_{i} (\vec{\theta})}{\partial \theta_{j}} \bigg) \bigg]^{-1} \text{ , }
\end{equation}
where $\delta_{ij}$ is the Kronecker delta, and $\vec{\theta}$ is the angular position on the sky. 
The scaled deflection field in turn depends on the surface mass density $\Sigma$, 
\begin{equation}
    \vec{\alpha} (\vec{\xi}) = \frac{D_{ds}}{D_{s}} \frac{4G}{c^{2}} \int \Sigma (\vec{\xi'}) \frac{\vec{\xi}-\vec{\xi'}}{|\vec{\xi}-\vec{\xi'}|^{2}} d^{2} \xi' \textrm{ , }
\end{equation}
where $\vec{\xi}$ is the physical position on the lens plane, $D_{ds}$ is the angular diameter distance between the deflector (the cluster lens) and the source, and $D_{s}$ is the angular diameter distance between the observer (z=0) and the source. 
The (in)flexibility in the mass model is translated into that of the deflection field, which ultimately limits the accuracy of the modeled magnification. 

For parametric models, their flexibility is inherently limited because the parametric halos typically have few free parameters and thus have restricted shapes. 
For most free-form, or non-parametric models, however, flexibility is still somewhat limited by the particular choice of basis functions that one places on a grid \citep{diego05, liesenborgs06}. 
This is because the mass model is a summation of the basis functions that are required to be non-negative. 
A sum of basis functions with non-negative coefficients result in a model that can only be \textit{flatter} than one single basis function, but not steeper. 
The only truly non-parametric model is a grid of independent pixels, each having its own free parameter. 
Such a model indeed has unlimited flexibility, but the number of free parameters will be unfeasibly large for any practical spatial resolution. 
These limitations found in the current landscape of cluster lens modeling call for a new approach. 

In this paper, I propose a new free-form parametrization that can model mass distributions of galaxy clusters with high degrees of accuracy, efficiency, and flexibility.  
This approach is inspired by the widely-used JPEG image compression algorithm, which uses a linear combination of two-dimensional cosine functions to model the original image.  
The fact that JPEG image compression is able to accurately and efficiently approximate all kinds of images makes it an interesting case to consider for modeling the mass distributions within galaxy clusters. 

This paper is structured as follows. 
In section 2, I explain the core idea of JPEG image compression that is relevant to lens modeling. 
Next, I describe the lens parametrization, and apply it to model an N-body simulated cluster first using all the pixels of the deflection field as constraint, and then using only 200 random pixels. 
Section 3 presents the best-fit deflection field models and the reconstructed mass and magnification models. 
In section 4 I discuss the strengths and improvement in efficiency of the JPEG parametrization compared to existing free-form lens modeling methods. 
In section 5, I provide a summary of the methodology and results.

\section{Methodology}

\subsection{Utility of JPEG image compression\\for cluster lens modeling}

I begin by describing the utility of JPEG image for encapsulating information in an efficient way. 
Without dwelling on every detail of how it works, this section provides a brief description that is relevant to cluster lens modeling. 

Depending on the compression ratio, JPEG images can be extremely accurate approximations of their original, uncompressed counterparts. 
They are often visually indistinguishable from the original images while taking up only 10 to 20 times less storage space. 

\begin{figure*}
    \centering
    \includegraphics[width=18cm]{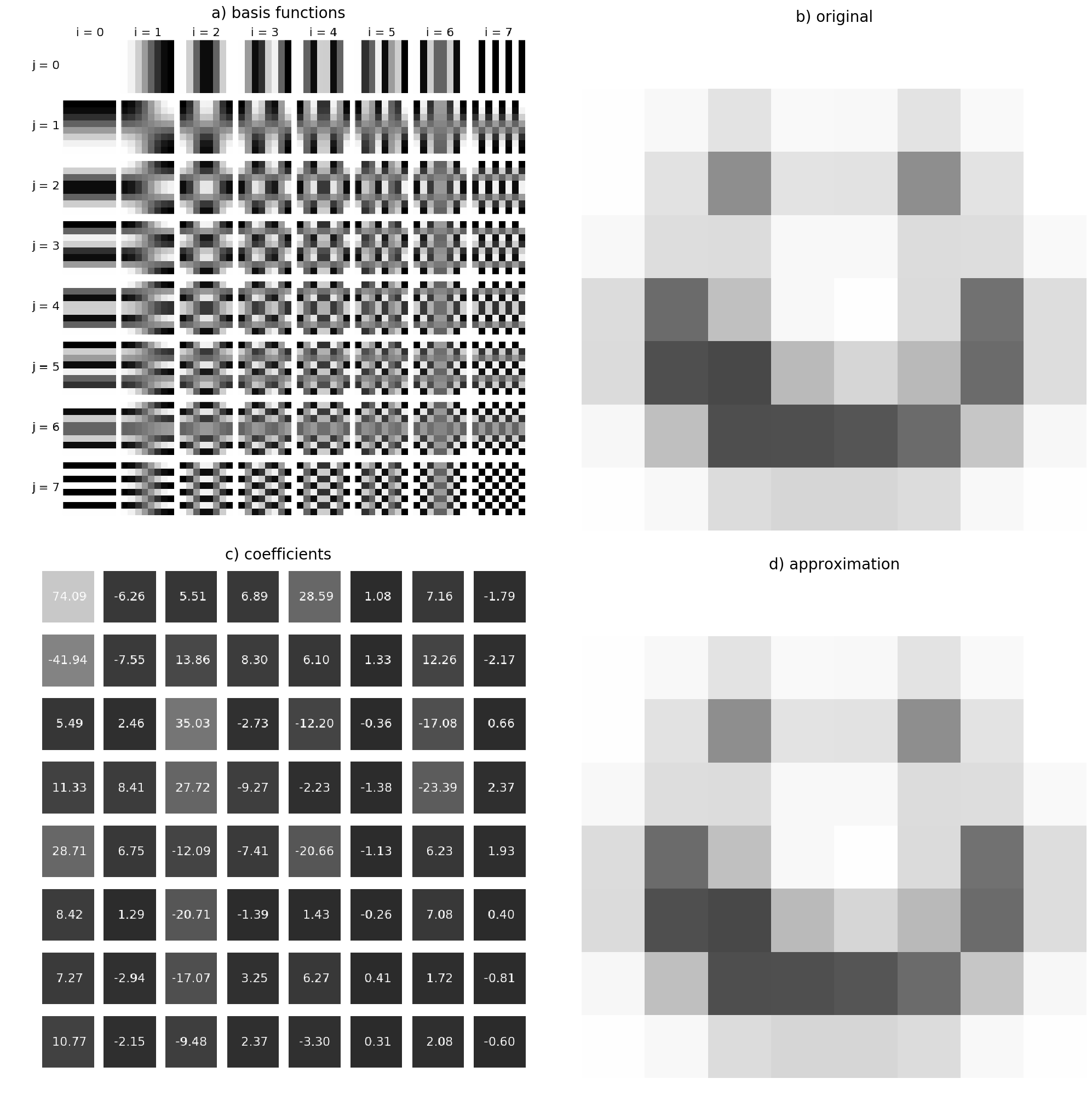}
    \caption{A brief illustration of the JPEG image compression method. 
    \textit{a)}: The 64 basis functions used in JPEG image compression. 
    Each basis function is the product of a two-dimensional cosine function varying in the x-direction and one in the y-direction. 
    Products of short-wavelength functions are useful at capturing small-scale features in the image, and vice versa. 
    Each basis function is 8 pixels by 8 pixels across. 
    \textit{b)}: The original image of a Gaussian-smoothed smiley face. 
    It has a size of 8 pixels by 8 pixels. 
    \textit{c)}: The optimal coefficients of the basis functions such that the original image is best reproduced.
    The background lightness of each sub-panel denotes the absolute amplitude of that coefficient. 
    \textit{d)}: The approximate image is the linear combination of all the basis functions weighted with the coefficients listed in \textit{c)}. 
    In this case, the replication is perfect because the number of parameters has not changed. 
    In real-world applications, short-wavelength components can often be discarded to save storage space without significantly degrading image quality. }
    \label{fig:dct}
\end{figure*}

The original, uncompressed image is first divided into 8$\times$8 pixel blocks. 
Each of them is approximated separately using the same treatment. 
A color image consists of a luminosity and a color component. 
Here, only the approximation for the former is used as an example. 
Panel $b)$ of figure \ref{fig:dct} shows an 8 pixel $\times$ 8 pixel image of a hand-drawn figure smoothed with a Gaussian filter. 

At the heart of JPEG compression is a set of 8 $\times$ 8 = 64 cosine basis functions, 
\begin{equation}\label{eq:jpeg_bf}
    g_{i, j}(x, y) = \mathrm{cos}(\frac{ix}{2}) \times \mathrm{cos}(\frac{jy}{2})
    \text{ , }
\end{equation}
where $i$ and $j$ are integers ranging from 0 to 7, while $x$ and $y$ are pixel coordinates and each take 8 equally separated values from 0 to 2$\pi$. 
14 of the basis functions ($j=0$ and $i > 0$, and $i = 0$ and $j > 0$) consist of two-dimensional cosine functions of various wavelengths that propagate in the x- and y- directions, respectively. 
49 of them ($i > 0$ and $j > 0$) are the products of each of these cosine basis functions oscillating in the x-direction with each of those oscillating in the y-direction. 
The remaining one ($i = j = 0$) is constant over all pixels and simply act as a `floor'. 
Panel $a)$ of Figure \ref{fig:dct} shows what the basis functions look like. 

As with most image files, the input image has pixel values ranging from 0 to 255.
However, the basis functions range from -1 to +1. 
Therefore, to approximate the image with those basis functions, it has to be centered around 0 by subtracting 128 throughout. 
In the end, all pixel values are raised by 128 to give the final output. 

Each basis function has its own coefficient, which can be negative or positive. 
They are calculated such that the linear combination of all the basis functions gives the best approximation. 
This technique central to JPEG image compression is called `discrete cosine transform'. 
In our example, we found that the original image is best reproduced with the particular set of coefficients listed in panel $c)$ of Figure \ref{fig:dct}. 
In fact, the original image is perfectly reproduced after rounding the pixel values to integers. 

At this point, the approximate image actually uses the same amount of information as the original image (64 coefficients versus 64 pixel values). 
In real-world applications, short-wavelength components can often be discarded without degrading the approximate image much, thus reducing the storage space needed. 
As one can see, the short-wavelength coefficients are typically smaller in Figure \ref{fig:dct} $c)$.  
In addition, the human eye is even less sensitive to color changes than to brightness changes. 
Therefore, compression in color space can be even more aggressive than in luminosity without making much noticeable difference. 

In this subsection, I have briefly demonstrated how JPEG compression works. 
For the presented case, the number of free parameters (coefficients), even at high compression, is of the order of the number of pixels. 
By contrast, in the setting of cluster lens modeling, which I will introduce in the next subsection, a usable lens model should be at least a few hundred pixels across, which means the number of pixels is of the order of a hundred thousand. 

Since, fortunately, abrupt pixel-to-pixel variations in general are not common in cluster lenses, the number of free parameters can be set drastically lower than the number of pixels, at the expense of ensuring that the information on the smalleset spatial scale is correct. 
Ideally, the number of free parameters should roughly match that of the constraints available, which is about a few hundred for a well-studied Frontier Field-like cluster including weak lensing constraints. 

\subsection{Example simulated cluster}

\begin{figure*}
    \centering
    \includegraphics[width=18cm]{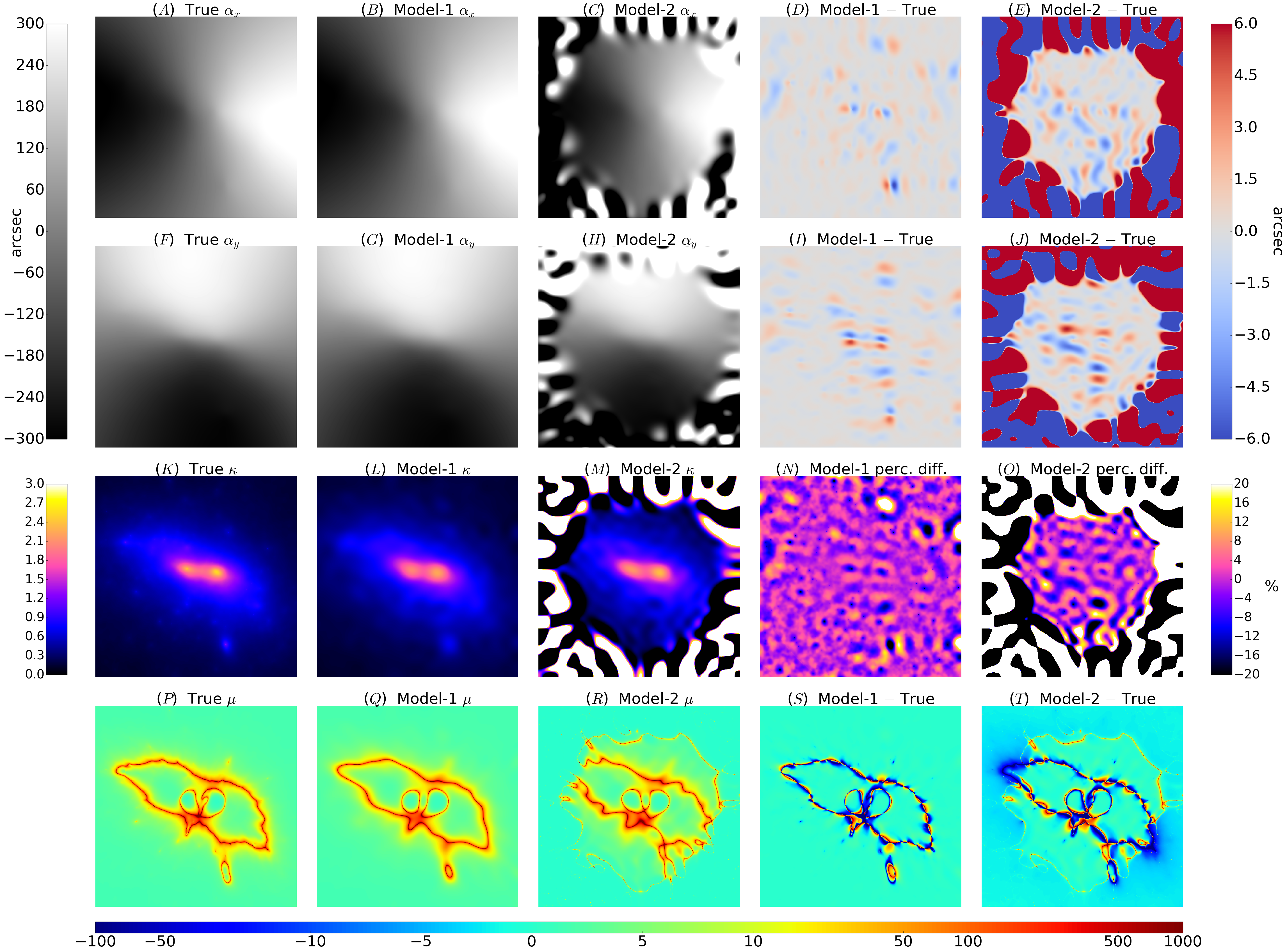}
    \caption{
    Comparing the true deflection field ($\alpha_{x}$, $\alpha_{y}$), convergence $\kappa$, and magnification $\mu$ of an N-body simulated cluster, \textit{Hera}, with those of Model-1 and Model-2 representations. 
    The first column shows the true quantities; the second shows the best-fit model constructed by minimizing the total difference between the true and model deflection field; the third shows the best-fit model constructed by minimizing the difference in deflection at only 200 randomly sampled locations; the fourth and fifth columns highlight the differences between the models and the truth. 
    The first best-fit model is constructed by minimizing the total difference in deflection (second column). 
    It demonstrates that the parametrization with cosine basis functions is indeed flexible enough to accurately capture most details of the lens. 
    In terms of deflection, the discrepancy (panels ($D$) and ($I$)) is typically less than 1" with the exception for regions immediately next to the core regions of galaxy halos, where the discrepancy can reach 6". 
    This is because the cosine basis functions are unable to capture any details smaller than half the shortest wavelength, which is 7" in this case. 
    This same problem is evident in panel ($N$), which shows the percentage difference between the true (panel $K$) and the model convergence (panel $L$), where all the core regions of galaxy halos are under-estimated by $>$10\%. 
    The global mass model, however, is accurate to within a few percent. 
    The true magnification (panel $P$) is also mostly accurately captured by the model (panel $Q$), with the exception of small-scale fluctuations. 
    There is neither significant systematic discrepancies in the location of the critical curves nor over large-scale, low-magnification regions.
    The second best-fit model is constructed by minimizing the difference in deflection at only 200 locations. 
    The positions are randomly and uniformly distributed within a central circular region to imitate a sampled constraint of the deflection field with multiply lensed galaxies. 
    The discrepancies between this sampled model and the truth is slightly larger in the constraint region but still retains most of the features. 
    Beyond the constrained region, the best-fit model shows large fluctuations, as smoothness is not a criterion implemented in the optimization process. 
    The color bar for magnification at the bottom is roughly linear between $-$10 and $+$10, and is logarithmic beyond this range. 
    All quantities are calculated at a source redshift of $z=9$. 
    Each panel spans a field of view of 2.22'$\times$2.22'. }
    \label{fig:comparison}
\end{figure*}

To illustrate how a JPEG parametrization would work with a cluster mass distribution, I use as an example the lensing cluster \textit{Hera}, which is identified in a collisionless N-body cosmological simulation \citep{planelles14}. 
It is used in \citet{meneghetti17} to compare how well various lens modeling methods work. 
The cluster is located at a redshift of $z=0.507$ and has a total mass of $M = 9.4 \times 10^{14} h^{-1} \textrm{M}_{\odot}$. 
Its virial region is well-resolved with $\approx$10 million dark matter particles. 
For this study, the central 2.22'$\times$2.22' region of its surface mass density is rebinned to 500 $\times$ 500 pixels. 
Panel ($K$) in Figure \ref{fig:comparison} shows its dimensionless surface mass density (convergence $\kappa$) at $z=9$. 
The scaled deflection field (panels ($A$) and ($F$)) is then computed pixel-by-pixel. 

\subsection{Parametrization of cluster mass distribution}

In this study, I will experiment with modeling the surface mass density of clusters using 20 $\times$ 20 = 400 basis functions, which have the following form, 
\begin{equation}
    g_{i, j}(x, y) = \bigg[ \mathrm{cos}(\frac{ix}{2}) \times \mathrm{cos}(\frac{jy}{2}) \bigg] + 1
    \text{ , }
\end{equation}
where $i$ and $j$ range from 0 to 19, and $x$ and $y$ range from 0 to 2$\pi$. 
The basis functions are all raised by $+1$ relative to those in equation \ref{eq:jpeg_bf} because their summation represents the surface mass density, which has to be non-negative. 
The coefficients can be positive or negative - this freedom allows greater flexibility than existing free-form lens parametrizations do. 

\subsubsection{Fitting the entire deflection field: Model-1}

The full set of coefficients that best represents the surface mass density is determined in an iterative process (referred to as optimization thereafter). 
I derive it by fitting to the deflection field since in practice, that is what being directly constrained by strong lensing observations (other observational constraints such as weak lensing shear and relative magnification among multiple images are related to \textit{derivatives} of the deflection field). 
The total absolute difference between the true and the model deflection field is minimized using the \texttt{SLSQP} algorithm in \texttt{scipy.minimize} with all initial coefficients equal zero. 
The resultant best-fit model is denoted as Model-1. 

To speed up the optimization of the lens model, instead of integrating the deflection field from the mass model at each step, the deflection field of each of these basis functions are computed in advance, and are simply weighted by their coefficients and then summed to produce the overall deflection field model at each step. 
Note that, unlike the models constructed in \cite{meneghetti17}, this model is not derived by trying to match mock observables based on a cluster mass distribution. 
Instead, the mass distribution is determined by directly fitting the deflection field. 
Therefore, the results for Model-1 only demonstrate the best achievable outcome using a JPEG-like representation. 

\subsubsection{Fitting the sampled deflection field: Model-2}

While Model-1 is optimized over all 250,000 pixels, in practice, lensing constraints are available over a much smaller number of points within the overall high-magnification area.
As such, it is useful to consider a representation which is more consistent with what one would derive from actual lensing constraints. 
For this reason, I optimize the deflection field model for Hera at only 200 sample points, which roughly equals the number of multiple images in a well-observed cluster. 
This best-fit model is denoted as Model-2. 
Like Model-1, Model-2 also consists of 400 basis functions. 
The 200 points of constraint are randomly and uniformly distributed within a central circular region of 7.2' radius. 
The placement of these sampling points cover most of the critical curves at $z=9$, and approximates the spatial distribution of multiple images (including radial images) of sources over a range of redshifts. 

\section{Results}

\subsection{Accuracy and precision of Model-1}

Panels ($B$) and ($G$) in Figure \ref{fig:comparison} show the horizontal and vertical components of the best-fit deflection field model at $z=9$, respectively. 
All the features of the deflection field are accurately reproduced with the exception of those near high-density, compact halos. 
The differences between the true and model deflection field can be seen more clearly in panels ($D$) and ($I$), where the contrast is increased by 50 times and color-coded. 
In regions where multiple images are present, the difference is in general within 1". 
The difference is largest ($\approx$6") near the central regions of the most massive halos. 
These discrepancies arise because small-scale features such as the cores of galaxy halos cannot be accurately reproduced by even the basis functions with the shortest `wavelengths'. 
In this case, one half of the shortest wavelength equals 7". 
Any structures smaller than this length scale cannot be modelled. 
As a result, the magnitude of the deflection is underestimated near the high-density, compact regions. 

The same problem is evident from panel ($N$) of Figure \ref{fig:comparison}, which shows the percentage difference between the true and best-fit convergence at $z=9$. 
The true convergence is in general reproduced to within a few percent, but can be underestimated by $>$10\% at the cores of galaxy halos. 
Some local over-estimation around an isolated halo towards the bottom of the field-of-view is caused by the inability to reproduce highly elliptical galaxy halos given the current number of basis functions (a larger number of basis functions will better reproduce small-scale features, including ellipticity). 

\begin{figure*}
    \centering
    \includegraphics[width=18cm]{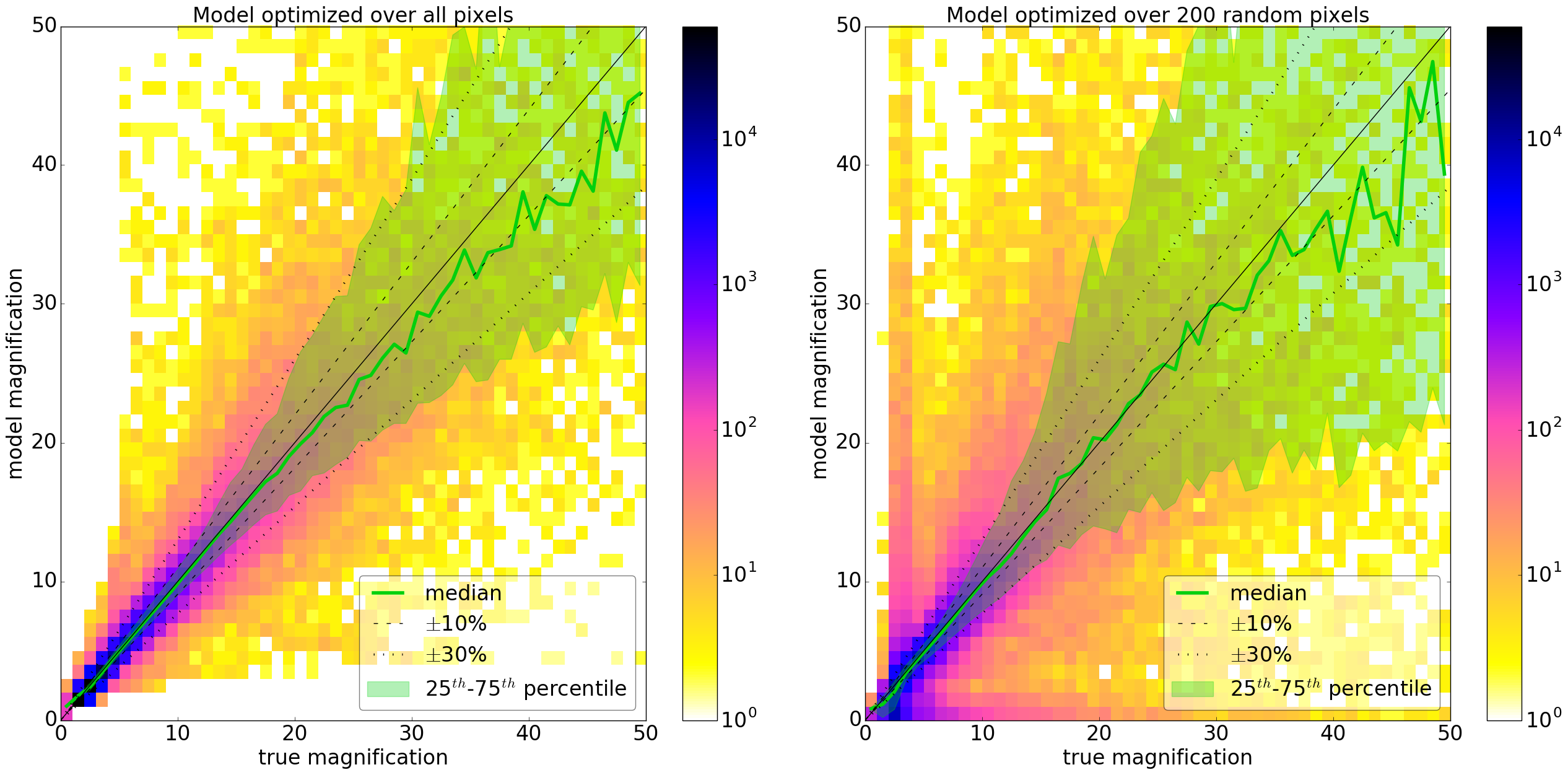}
    \caption{
    Model versus true magnification. 
    The colored surface density plot shows the distribution of pixels. 
    \textit{Left}: the best-fit model constructed by minimizing the total difference between the true and model deflection field (Model-1). 
    \textit{Right}: the best-fit model constructed by minimizing the difference in deflection at 200 random locations (Model-2). 
    Only the constrained region is shown for Model-2. 
    The green line shows the median model magnification as a function of true magnification. 
    The green shaded region shows the distribution within the 25th and 75th percentiles. 
    The solid black line is the one-to-one line. 
    The dashed and the dotted black lines denote $\pm10\%$ and $\pm30\%$ deviation from the true magnification, respectively. 
    The median magnification of Model-1 shows a slight underestimation beyond a true magnification of 20 but mostly remains within 10\% accurate all the way out to 40. 
    The precision (green shaded area) is within 10\% below a true magnification of 10, and within 30\% below 30. 
    With much fewer constraints, the median of Model-2 is similarly accurate to 10\% in the median up to 40 times magnification. 
    However, the precision degrades much more quickly - the 25$^{th}$-75$^{th}$ percentile exceeds 10\% of the true magnification beyond a magnification of 5, and 30\% beyond a magnification of 12. 
    This plot is made following figures 23 and 24 in \cite{meneghetti17} for convenient comparison. }
    \label{fig:mu_diff}
\end{figure*}

The magnification map is accurately reproduced except the smallest features, as shown in panels ($P$), ($Q$), and ($S$) of Figure \ref{fig:comparison}. 
There is neither significant systematic offsets in the location of the critical curves nor in large-scale, low-magnification regions. 
Figure \ref{fig:mu_diff} plots, pixel by pixel, the model magnification versus the true magnification. 
The median model magnification is in general accurate to within 10\% to $\mu_{true}=40$ with only slight systematic deviation of a few percent beyond $\mu_{true}>20$. 
Following the convention of \citet{meneghetti17}, precision is defined as the $25^{th}$ and $75^{th}$ percentiles of the distribution of $\mu-\mu_{true}$, which is within 10\% up to $\mu_{true}=10$ and 30\% up to $\mu_{true}=30$. 

\subsection{Accuracy and precision of Model-2}

Figure \ref{fig:comparison} also compares the true deflection field, convergence, and magnification with those of Model-2. 
It shows larger differences from the truth than the one optimized using all pixels, but still recovers most of the prominent features within the constrained region. 
Beyond where constraints exist, the best-fit model shows large fluctuations as smoothness is not a criterion in the optimization process. 

The right panel of Figure \ref{fig:mu_diff} shows the comparison between the model and the true magnification for pixels within the constrained region. 
Model-2 shows a larger dispersion compared to Model-1. 
The median model magnification is still accurate to within 10\% up to 40$\times$ magnification. 
However, the precision of Model-2 degrades much more quickly - the 25$^{th}$-75$^{th}$ percentile exceeds 10\% and 30\% difference at magnification of 5$\times$ and 12$\times$, respectively.

\section{Discussion}

What has been demonstrated in this paper is simply a direct fit of a simulated surface mass density with the cosine basis functions. 
By no means is it a working lens modeling code. 
Nevertheless, it shows the capability of this `JPEG' approach, which so far appears to be a promising choice of parametrization for free-form lens modeling. 
The biggest hurdle needed to be overcome in order to make this proof-of-concept into a functioning lens modeling code is to optimize the lens model with lensing constraints. 
This will be described in a future paper (Lam et al., in prep.). 

\subsection{Comparison with other free-form Hera models}

A significant point of this study is to present an approach that efficiently encapsulates information about the surface mass density of a galaxy cluster. 
Therefore, it is useful to compare our approach to others used in the literature. 
The present JPEG representation of the Hera cluster makes use of 400 free parameters. 

Here I compare it with the six free-form models from \citet{meneghetti17}. 
In that study, a number of lens modeling methods are benchmarked by reconstructing lens models for \textit{Hera} using simulated lensing constraints. 
19 background galaxies are strongly lensed into 65 multiple images, and additional background galaxies are weakly lensed with a surface number density of $\approx$14 arcmin$^{-2}$. 

The Bradac-Hoag model, constructed using the method SWUnited \citep{bradac05, bradac09}, has 5497 free parameters. 
These free parameters correspond to values in the gravitational potential over the same number of points. 
These points are distributed over an iterative, multi-resolution grid which is coarse in the cluster outskirts where constraints are sparse, and fine in the cluster center, where the brightest cluster galaxies and most multiple images reside. 
In addition to the 65 multiple images, this model is also constrained by 2102 weak lensing ellipticity measurements. 

Two out of three variants of Diego models, and the Lam model, constructed using the method WSLAP+ \citep{diego05, diego07, sendra14}, have 1027 free parameters. 
WSLAP+ models the mass distribution primarily as the summation of a grid of two-dimensional Gaussian functions. 
The two Diego models and the Lam model aforementioned are evaluated on regular grids. 
The third variant of the Diego models is evaluated on an adaptive grid and has only 304 free parameters. 
The grid of the adaptive Diego model has varying resolution over the field-of-view. 
The resolution of the grid is iteratively increased over regions where the previous optimization step found more mass compared to other regions. 
The adaptive grid allows the model to perform comparably with regular-grid models using roughly only half the number of free parameters. 
A second component of WSLAP+ accounts for small-scale mass features, that is, cluster galaxies. 
All three Diego models use the light profile of the cluster galaxies as the second component while the Lam model uses an NFW parametrization. 

The number of free parameters in the set of GRALE \citep{liesenborgs06} models ranges from 600 to a couple of thousands\footnote{The initial number of free parameters range from 600 to 1000, which increases during optimization. 
The exact final number of free parameters is not stated in \citet{meneghetti17}. }
The mass model comprise of a uniform mass sheet and a large number of projected Plummer spheres. 
In addition to using multiple images as constraints, the GRALE model is also constrained by ``null space", which are regions where no strongly lensed galaxies are present. 

In summary, the JPEG parametrization appears to be able to capture features at both small- and large-scales while using significantly fewer free parameters compared to most existing free-form lens modeling methods. 
Although the number of free parameters can be chosen arbitrarily, and that the JPEG models presented here are constructed in an entirely different way from the ones in \citet{meneghetti17}, a brief comparison nonetheless gives the reader a better idea of the potential gain in efficiency that the JPEG approach has.

\subsection{Progressive optimization}
Although not currently implemented, parametrizing the lens model as a linear combination of cosine basis functions allows convenient \textit{progressive} optimization of the lens model. 
While each of the two models takes 1.5 days to optimize, optimizing with lensing constraints will likely take much longer. 
The reason is that the chi-squared `landscape' in the parameter space is likely to be much more complex with lensing constraints, thus the optimization algorithm will take longer to find the next direction of descent. 

Progressive optimization mitigates this problem. 
To begin, the lens model is only parametrized with a handful of long-wavelength basis functions. 
Optimizing a small number of free parameters will quickly yield a lens model that is roughly correct on the large scale. 
These optimized coefficients can then be used as the initial solution for the next optimization, which adds a few shorter-wavelength basis functions to the model. 
Using this strategy, the optimization algorithm will start exploring the parameter space closer to the global minimum compared to the case where it starts at some other arbitrary or random locations, thus shortening the time required to reach convergence.

\section{Summary}

In this paper, I propose a new approach to free-form cluster lens modeling inspired by the parametrization used in JPEG image compression. 
Unlike the conventional approach to free-form cluster lens modeling, which places two-dimensional basis functions on a grid, this approach uses orthogonal cosine functions of different wavelengths and their products as basis functions. 

This approach has the advantage of offering greater flexibility in the lens model while using fewer free parameters. 
As explained in section \ref{sec:intro}, flexibility is key to obtaining accurate magnification, which many other science goals rely on, such as the exciting prospect of detecting faint-end turn-overs in the high-redshift galaxy luminosity functions.

As a proof of concept, I demonstrate the capability of this approach by directly fitting the deflection fields of 400 cosine basis functions to that of an N-body simulated cluster, \textit{Hera} \citep{meneghetti17}.  
For a well-studied, Frontier Field-like cluster, this amount of free parameters roughly matches that of available constraints (including weak lensing constraints). 
The overall morphology of the mock lens is accurately reproduced except for the central regions of individual halos. 
This is because the cosine basis functions are unable to capture any details smaller than half of their shortest wavelength. 

The accuracy of the best-fit model is quantified by the median model magnification, which varies within 10\% of the true magnification out to a magnification of 40. 
The precision is quantified by the 25$^{th}$ to 75$^{th}$ percentile, which is within 10\% up to $\mu_{true}=10$ and 30\% up to $\mu_{true}=30$. 

A second, more realistic model is obtained by fitting the basis functions to a sample of only 200 positions drawn randomly from the strong lensing region and of use for demonstrating use of the `JPEG approach' in a more realistic setting. 
It has a similar accuracy within the constrained region although precision degrades more quickly - the 25$^{th}$ to 75$^{th}$ percentile exceeds 10\% and 30\% at 5 and 12 times magnification, respectively. 
These encouraging capabilities show that this `JPEG lensing' approach could be a promising technique in future cluster lens modeling. 

In the future, I plan to reconstruct the simulated N-body cluster, \textit{Hera}, with the JPEG parametrization using multiple images and weak lensing ellipticity measurements as constraints. 
Once that is demonstrated, a more streamlined progressive optimization will be implemented, which will potentially result in faster optimization and more accurate models.

\acknowledgments
The author thanks Rychard Bouwens and Jose Diego for valuable feedback on both the write-up and science of the manuscript, Jori Liesenborgs, Liliya Wiliams, Dan Coe, and Masamune Oguri for stimulating discussions, Massimo Meneghetti for coordinating the HFF comparison project, and the HFF community for contributing to the conversation with all their collective efforts.

\end{document}